\documentclass[twocolumn,preprintnumbers,amsmath,amssymb]{revtex4}
\usepackage{cases}
\usepackage{amssymb}
\usepackage{txfonts}
\usepackage{graphicx}
\usepackage{dcolumn}
\usepackage{bm}
\usepackage{amssymb}
\usepackage{amsmath}
\usepackage{epsfig}
\usepackage{subfigure}
\usepackage{overpic}
\begin{document}
\title{All-optical transistor with cavity polaritons}
\author{Zi-Fa Yu$^{1,2}$}
\email[Corresponding author. Email: ]{Yuzifa@nwnu.edu.cn}
\author{Ju-Kui Xue$^{2}$}
%\email[Corresponding author. Email: ]{xuejk@nwnu.edu.cn}
\author{Jinkui Zhao$^{1,3}$}
%\email[Corresponding author. Email: ]{jkzhao@iphy.ac.cn}
\author{Wu-Ming Liu$^{1,3}$}
%\email[Corresponding author. Email: ]{wliu@iphy.ac.cn}
\address{$^{1}$Beijing National Laboratory for Condensed Matter Physics, Institute of Physics, Chinese Academy of Sciences, Beijing 100190, China}
\address{$^{2}$Key Laboratory of Atomic $\&$ Molecular Physics and Functional Materials of Gansu Province, College of Physics and Electronic Engineering, Northwest Normal University, Lanzhou, 730070, China}
\address{$^{3}$Songshan Lake Materials Laboratory, Dongguan 523808, China.}
\begin{abstract}
we investigate the transmission of probe laser beam in a coupled-cavity system with polaritons by using standard input-output relation of optical fields, and proposed a theoretical schema for realizing a polariton-based photonic transistor. On account of effects of exciton-photon coupling and single-photon optomechanical coupling, a probe laser field can be either amplified or attenuated by another pump laser field when it passes through a coupled-cavity system with polaritons. The Stokes and anti-Stokes scattered effect of output prober laser can also be modulated. Our results open up exciting possibilities for designing photonic transistors.
\end{abstract}

 \maketitle

\section{Introduction}
An optical transistor is a device which uses photons as signal carriers to control the probe laser field by using another pump laser field \cite{transistor1,transistor2}. In comparison to the electronic transistor, the optical transistor possesses much higher transfer rates \cite{transfer}, causes much less hardware heating \cite{heat}, and remains effective at the nanometer scale \cite{nanometerscale}. Therefore, the optical transistor is often heralded as the next step of quantum information processing, and is the basis of optical computation and communication \cite{nextstep}. However, there are some challenges for building a such optical transistor because photons rarely interact each other and photon-photon interactions are very weak even in nonlinear optical media \cite{weakinteract}. To remedy this shortcoming, several schemes for mediating weak light beams are proposed, such as enhancing resonances in optical emitters \cite{emitters1,emitters2,emitters3} and achieving strong coupling between light and matter \cite{Lightmatter1,Lightmatter2,Lightmatter3}. A tightly focused laser beam can be coherently modulated by another gating beam via individual optical emitters realized by embedding a quantum dot in a photonic crystal nanocavity \cite{nanometerscale}, embedding dye molecules in organic crystalline matrices \cite{transistor2} and embedding nitrogen vacancies in suitable host materials \cite{hostmat}. Light-matter interactions have been presented in a cavity optomechanical system \cite{optocavity1,optocavity2,optocavity3,optocavity4,optocavity5}, where individual atoms, ultra-cold atoms, or Bose-Einstein condensates are coupled to photons. With such strong coupling interactions, the mechanical motion of matters can easily modulate the transmission of photons \cite{backfeed1,backfeed2}, creating sidebands below (the Stokes sideband) and above (the anti-Stokes sideband) the drive frequency, as well as leading to the effect of electromagnetically induced transparency \cite{ETC1}.

Cavity optomechanical system is essentially a hybrid physical field of optical resonators coupled to mechanical oscillators \cite{hybrid1,hybrid2,hybrid3,hybrid4}. In such a system, the radiative pressure and dynamical backaction enable the optical control of mechanical oscillators, which also reacts on optical resonators. Recently, a new type of cavity optomechanical system with the strong light-matter coupling between photons and excitons have been reported in a GaAs/AlAs quantum-well microcavity experimentally \cite{Fainstein2013}. Such light-matter coupling results in the emergence of new quasiparticles, i.e., exciton polaritons. This semiconductor optomechanical system has its own advantages over atomic optomechanical systems. Exciton polaritons obey bosonic statistics, and can be condensed at room temperatures owing to extremely light effective mass \cite{copy3,copy4,copy5,copy6,copy7}. Exciton-polariton condensates possess the quantum nonequilibrium and non-Hermitian nature \cite{copy10,copy11,copy12,copy13}. They can be directly imaged in momentum and real spaces through the cavity photoluminescence in experiments \cite{copy8,copy9}. Hence, exciton-polariton condensates have not only fascinations in fundamental investigations but also potential applications in functional polariton devices, as well as have aroused enormous attentions both in theories and experiments. Moreover, in such an optomechanical system based on exciton polaritons, the strong coupling effect between excitons and photons can modulate the transmission of photons, and lead to novel optomechanically induced transparency, which is useful for realizing optical transistors.
%%%%%%%%%%%%%%%%%%%%%%%%%%%%%%%%%%%%%%%%%%%%%%%%%%%%%%%%%%%%%%%%%%%
\begin{figure}[htpb]
\begin{center}
\includegraphics[width=8.0cm]{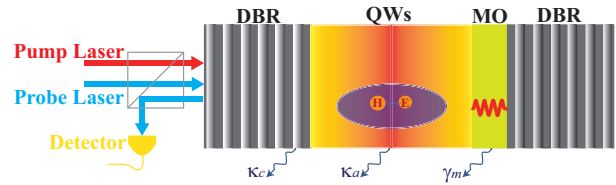}
\end{center}
\caption{Sketch of an optomechanical device in the simultaneous presence of a pump laser and a probe laser. The probe beam transmitted from the device are monitored by a detector. Optical cavity and mechanical oscillator (MO) is formed by two distributed Bragg reflectors (DBR) containing with quantum wells (QWs) placed at the antinode of the cavity field. There is a strong coupling effects between cavity photons and excitons in QWs.}\label{fig1}
\end{figure}
%%%%%%%%%%%%%%%%%%%%%%%%%%%%%%%%%%%%%%%%%%%%%%%%%%%%%%%%%%%%%%%%%%%%%

In this paper, we investigate the transmission of probe laser beam in an optomechanical resonator device with strong exciton-photon coupling (Fig. \ref{fig1}), where optical cavity and mechanical oscillator is formed by two movable end distributed Bragg reflectors, containing with quantum wells placed at the antinode of the cavity field. The cavity is driven by a strong pump laser with frequency $\omega_{pu}$ and a weak probe laser with frequency $\omega_{pr}$. This strong optical-mechanical coupling has been experimentally realized in a vertical GaAs/AlAs microcavity \cite{Fainstein2013}. Using standard input-output relation \cite{IOR}, we calculate the probe transmission coefficient and analyze transmission spectrum. The result indicates that a probe laser field can be either amplified or attenuated by suitably adjusting another pump laser field in a cavity optomechanical system with exciton-photon coupling. Moreover, we can also modulate the Stokes and anti-Stokes scattered effect of output prober laser. Our results provide a theoretical schema for all-optical transistor based on a coupled polariton cavity system.

\section{Model}
Motivated by the experiment in Ref. \cite{Fainstein2013}, We consider an optomechanical resonator device as shown in Fig. \ref{fig1}, where optical cavity and mechanical oscillator is formed by two movable end distributed Bragg reflectors, containing with quantum wells placed at the antinode of the cavity field. The cavity is driven by a strong pump laser with frequency $\omega_{pu}$ and a weak probe laser with frequency $\omega_{pr}$. In the frame rotating at the pump laser frequency $\omega_{pu}$, one can obtain the generic Hamiltonian of the system \cite{Fainstein2013,EPCM1,EPCM2}:
\begin{equation}\label{eq1}
\begin{split}
&\hat{H}=\hbar\omega_{m}\hat{b}^{\dag}\hat{b}-\hbar\Delta_{a}\hat{a}^{\dag}\hat{a}-\hbar\Delta_{c}\hat{c}^{\dag}\hat{c}
+\hbar\Omega(\hat{a}^{\dag}\hat{c}+\hat{c}^{\dag}\hat{a})
\\&+\hbar g\hat{c}^{\dag}\hat{c}(\hat{b}^{\dag}+\hat{b})
\!+\!{\rm i}\hbar[(\varepsilon_{pu}\!+\!\varepsilon_{pr}{\rm e}^{-{\rm i}\delta t})\hat{c}^{\dag}\!-\!(\varepsilon_{pu}\!+\!\varepsilon_{pr}{\rm e}^{{\rm i}\delta t})\hat{c}],
\end{split}
\end{equation}
where $\hat{a}$, $\hat{b}$, and $\hat{c}$ correspond to field operators for quantum well excitons, mechanical oscillators, and cavity photons with resonance frequency $\omega_{a}$, $\omega_{m}$, and $\omega_{c}$, respectively. $\Delta_{a}=\omega_{a}-\omega_{pu}$ and $\Delta_{c}=\omega_{c}-\omega_{pu}$ are the detunings of the exciton mode and the cavity mode, respectively. $\Omega$ is Rabi coupling strength which describes the coupling effect between cavity photons and excitons in quantum wells. $g$ is the
single-photon optomechanical coupling strength which describes the coupling effect between cavity photons and mechanical oscillators. $\varepsilon_{pu}=\sqrt{\kappa_{c}P_{pu}/\hbar\omega_{pu}}$ ($\varepsilon_{pr}=\sqrt{\kappa_{c}P_{pr}/\hbar\omega_{pr}}$) represents the pump (probe) laser intensity with input power $P_{pu}$ ($P_{pr}$) and cavity photon decay rate $\kappa_{c}$. $\delta=\omega_{pr}-\omega_{pu}$ is the probe-pump detuning.

In order to deal with the mean response of optomechanical oscillators to probe laser field, we introduce the position-like operator of the Bogoliubov mode $\hat{Q}=(\hat{b}^{\dag}+\hat{b})/\sqrt{2}$ and momentum-like operator $\hat{P}={\rm i}(\hat{b}^{\dag}-\hat{b})/\sqrt{2}$, then consider nonlinear Heisenberg-Langevin equations $\partial_{t}\hat{O}=({\rm i}/\hbar)[\hat{H},\hat{O}]+\hat{\mathcal{N}}$, where $\hat{O}=\{\hat{Q},\hat{P},\hat{a},\hat{c}\}$, $\mathcal{N}$ is the corresponding input vacuum noise operator. Thus, upon the communication relations $[\hat{a},\hat{a}^{\dag}]=1$, $[\hat{c},\hat{c}^{\dag}]=1$, and $[\hat{P},\hat{Q}]={\rm i}$, the temporal evolutions of mechanical oscillator mode, quantum well exciton mode, and cavity photon mode can be obtained by following equations:
\begin{equation}\label{eq2}
\begin{split}
\ddot{Q}+&\gamma_{m}\dot{Q}+\omega_{m}^{2}Q=-\sqrt{2}g\omega_{m}c^{\dag}c-\sqrt{2\gamma_{m}}\dot{Q_{in}}(t),\\
\dot{a}=&({\rm i}\Delta_{a}-\kappa_{a})a-{\rm i}\Omega c-\sqrt{2\kappa_{a}}a_{in}(t),\\
\dot{c}=&({\rm i}\Delta_{c}-\kappa_{c})c-{\rm i}\Omega a-{\rm i}\sqrt{2}gQc
\\&+\varepsilon_{pu}+\varepsilon_{pr}{\rm e}^{-{\rm i}\delta t}-\sqrt{2\kappa_{c}}c_{in}(t),
\end{split}
\end{equation}
where, $Q\equiv\langle\hat{Q}\rangle$, $a\equiv\langle\hat{a}\rangle$, and $c\equiv\langle\hat{c}\rangle$. $\gamma_{m}$ and $\kappa_{c}$ are the decay rates of mechanical oscillator mode and exciton mode, respectively. $Q_{in}$, $a_{in}$ and $c_{in}$ are corresponding Markovian input noise operators with zero averages and delta correlation functions, under the high frequency condition, i.e., $\hbar\omega_{q}\gg k_{B}T$, where $q=\{a,c,m\}$, $T$ is the environmental temperature, and $k_{B}$ is the Boltzmann constant.

In the strong driving region, i.e., the probe laser intensity is much weaker than the pump laser intensity, the probe field can be treated as the perturbation of the steady state. Thus, we can linearize Eq. (\ref{eq2}) at the first order sidebands by using following ansatz:
\begin{equation}\label{eq3}
\begin{split}
&Q(t)=Q_{0}+Q_{+}{\rm e}^{-{\rm i}\delta t}+Q_{-}{\rm e}^{{\rm i}\delta t},\\
&a(t)=a_{0}+a_{+}{\rm e}^{-{\rm i}\delta t}+a_{-}{\rm e}^{{\rm i}\delta t},\\
&c(t)=c_{0}+c_{+}{\rm e}^{-{\rm i}\delta t}+c_{-}{\rm e}^{{\rm i}\delta t}.
\end{split}
\end{equation}
Upon inserting Eq. (\ref{eq3}) into Eq. (\ref{eq2}), one can obtain
\begin{equation}\label{eq4}
\begin{split}
&a_{0}=\Omega c_{0}/(\Delta_{a}+{\rm i}\kappa_{a}),\\
&a_{+}=\Omega c_{+}/(\Delta_{a}+\delta+{\rm i}\kappa_{a}),\\
&a_{-}=\Omega c_{-}/(\Delta_{a}-\delta+{\rm i}\kappa_{a}),
\end{split}
\end{equation}
\begin{equation}\label{eq5}
\begin{split}
&Q_{0}=-\sqrt{2}gc_{0}c_{0}^{*}/\omega_{m},\\
&Q_{+}=-\sqrt{2}g\omega_{m}(c_{0}^{*}c_{+}+c_{0}c_{-}^{*})/(-\delta^{2}-{\rm i}\gamma_{m}\delta+\omega_{m}^{2}),\\
&Q_{-}=-\sqrt{2}g\omega_{m}(c_{0}^{*}c_{-}+c_{0}c_{+}^{*})/(-\delta^{2}+{\rm i}\gamma_{m}\delta+\omega_{m}^{2}),
\end{split}
\end{equation}
and
\begin{equation}\label{eq6}
\begin{split}
&c_{0}=\frac{\varepsilon_{pu}}{(\kappa_{c}-{\rm i}\Delta_{c})+{\rm i}\sqrt{2}gQ_{0}+{\rm i}\Omega^{2}/(\Delta_{a}+{\rm i}\kappa_{a})},\\
&c_{+}=\frac{\varepsilon_{pr}-{\rm i}\sqrt{2}gc_{0}Q_{+}}{(\kappa_{c}-{\rm i}\Delta_{c}-{\rm i}\delta)+{\rm i}\sqrt{2}gQ_{0}+{\rm i}\Omega^{2}/(\Delta_{a}+\delta+{\rm i}\kappa_{a})},\\
&c_{-}=\frac{-{\rm i}\sqrt{2}gc_{0}Q_{-}}{(\kappa_{c}-{\rm i}\Delta_{c}+{\rm i}\delta)+{\rm i}\sqrt{2}gQ_{0}+{\rm i}\Omega^{2}/(\Delta_{a}-\delta+{\rm i}\kappa_{a})}.
\end{split}
\end{equation}
Solving Eqs. (\ref{eq4})-(\ref{eq6}), we can acquire
\begin{equation}\label{eq7}
\begin{split}
&\varepsilon_{pu}^{2}=n_{c}\left\{\left[\kappa_{c}^{2}+(\Delta_{\rm c}+2g^{2}n_{c}/\omega_{m})^{2}\right]\right.\\
&+\left.\left[2(\kappa_{a}\kappa_{c}-2\Delta_{a}g^{2}n_{c}/\omega_{m}-\Delta_{a}\Delta_{c})\Omega^{2}+\Omega^{4}\right]/(\kappa_{a}^{2}+\Delta_{a}^{2})\right\},
\end{split}
\end{equation}
and
\begin{widetext}
\begin{equation}\label{eq8}
c_{+}=\varepsilon_{pr}\left\{\frac{(\kappa_{a}^{2}+F^{2})[(\kappa_{c}-{\rm i}\delta)+{\rm i}(\Delta_{c}+C)]+\Omega^{2}(\kappa_{a}-{\rm i}F)}{(\kappa_{a}^{2}+F^{2})[(\kappa_{c}-{\rm i}\delta)^{2}+(\Delta_{c}+C)^{2}-D^{2}]+2\Omega^{2}[\kappa_{a}(\kappa_{c}-{\rm i}\delta)-F(\Delta_{c}+C)]+\Omega^{4}}\right\}.
\end{equation}
\end{widetext}
where, $n_{c}=|c_{0}|^{2}$, $A=2g^{2}/\omega_{m}^{2}$, $B=\omega_{m}^{2}/(\omega_{m}^{2}-{\rm i}\delta\gamma_{m}^{2}-\delta^{2})$, $C=A\omega_{m}n_{c}(1+B)$, $D=AB\omega_{m}n_{c}$, and $F=\Delta_{a}+\delta$.

To obtain the transmission spectrum of the probe field, we use standard
input-output relation, i.e., $c_{\rm out}(t)=c_{\rm in}(t)-\sqrt{2\kappa_{c}}c(t)$, where $c_{\rm out}$ and $c_{\rm in}$ are output and input optical field operators, respectively, which is appropriate and available for open cavity system. Here, $c_{\rm in}$ is determined by pump and probe laser fields. Therefore, we can obtain the output field in the following expression:
\begin{equation}
\begin{split}
&\langle c_{\rm out}(t)\rangle=(\varepsilon_{}/\sqrt{2\kappa_{c}}-\sqrt{2\kappa_{c}}c_{0}){\rm e}^{-{\rm i}\omega_{pu}t}\\
&+(\varepsilon_{pr}/\sqrt{2\kappa_{c}}-\sqrt{2\kappa_{c}}c_{+}){\rm e}^{-{\rm i}(\omega_{pu}+\delta)t}-\sqrt{2\kappa_{c}}c_{-}{\rm e}^{-{\rm i}(\omega_{pu}-\delta)t}.
\end{split}
\end{equation}
The total probe transmission coefficient is defined as the amplitude ratio of output and input optical fields at the frequency of probe laser field, i.e.,
\begin{equation}
T_{\rm pr}(\omega_{pr})=\frac{\varepsilon_{\rm pr}/\sqrt{2\kappa_{c}}-\sqrt{2\kappa_{c}}c_{+}}{\varepsilon_{\rm pr}/\sqrt{2\kappa_{c}}}=1-2\kappa_{c}\left[\frac{c_{+}}{\varepsilon_{\rm pr}}\right].
\end{equation}

Obviously, the transmission coefficient of probe laser beams can be effectively manipulated by pump laser field, which also depends on the coupling effect of cavity photons, quantum well excitons and mechanical oscillators. The property of transmission spectrum can be described by the amplitude and the phase of transmission coefficient $T_{\rm pr}(\omega_{pr})$. The amplitude can reveal amplifications and attenuations of probe transmission, while the phase is related to slow and fast probe transmission dynamics. Here, we focus on the control of probe field by using pump field, and define the amplitude $|T_{\rm pr}(\omega_{pr})|$ of probe transmission coefficient as probe transmission $T$ (i.e., $T=|T_{\rm pr}(\omega_{pr})|$) to indicate the amplification and the attenuation of probe field signals. Probe laser signals through the optomechanical system are amplified for $T>1$ and attenuated for $T<1$. Namely, the probe transmission can be controlled by appropriately adjusting pump laser intensity, probe-pump detuning, Rabi coupling strength, single-photon optomechanical coupling strength, exciton mode detuning and cavity mode detuning.

\section{Transmission spectrum of probe field}
\begin{figure}[htpb]
\begin{center}
\includegraphics[width=8.0cm]{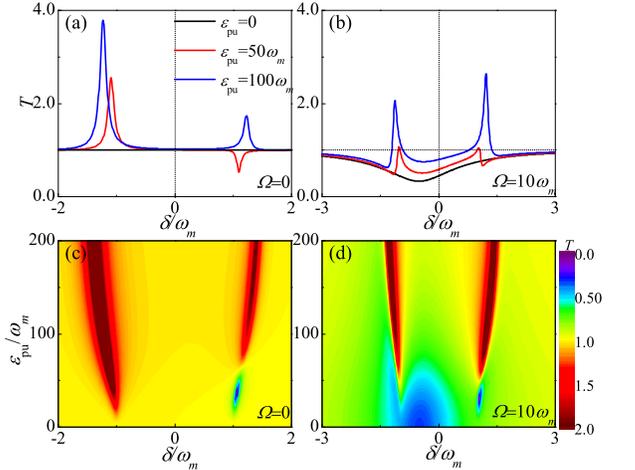}
\end{center}
\caption{Transmission $T$ of the probe beam for $g=10\omega_{m}$, $\Delta_{a}=\omega_{m}$, $\Delta_{c}=100\omega_{m}$, $\kappa_{a}=\omega_{m}$, $\kappa_{c}=100\omega_{m}$, and $\gamma_{m}=0.1\omega_{m}$. (a) and (b): Transmission $T$ versus probe-pump detuning $\delta$ for different pump laser intensity $\varepsilon_{pu}$. (c) and (d): Transmission $T$ in the $\delta-\varepsilon_{pu}$ plane.}\label{fig2}
\end{figure}

Transmission spectrum of probe field can be used to describe the property of output optical field. It can be effectively modulated by the pump laser field. In the absence of the coupling between cavity photons and quantum well excitons, i.e., the Rabi coupling strength $\Omega=0$, the output probe laser is modulated by cavity photons and mechanical oscillators. In this case, probe transmission coefficient $T_{\rm pr}(\omega_{pr})=-[\kappa_{c}+{\rm i}(\delta+\Delta_{c})]/[\kappa_{c}-{\rm i}(\delta+\Delta_{c})]$ without pump laser field, i.e., $\varepsilon_{pu}=0$, and we have $T=1.0$ which indicates the probe laser can completely transmit the optomechanical cavity (see black solid line of Fig. \ref{fig2}(a)). When a pump field is driven on the cavity, the probe transmission spectrum strongly depends on pump intensity and probe-pump detuning. For a weak pump intensity, the probe transmission is amplified at the resonant region with a red probe-pump detuning (i.e., $\delta<0$), and attenuated at the resonant region with a blue probe-pump detuning (i.e., $\delta>0$) as shown red solid line of Fig. \ref{fig2}(a). That is, a weak pump laser can induce the amplified Stokes light and the attenuated anti-Stokes light scattering from intracavity field. However, for a strong pump intensity, the probe transmissions are both amplified at the red and blue detuning resonant regions (see blue solid line of Fig. \ref{fig2}(a)). Therefore, a strong pump laser can generate the Stokes light and the anti-Stokes light which are both amplified. Moreover, as the pump intensity $\varepsilon_{pu}$ increases (Fig. \ref{fig2}(c)), the Stokes scattered light is gradually amplified and the corresponding resonant probe-pump detuning shifts red, while anti-Stokes scattered light is firstly attenuated then amplified and the corresponding probe-pump detuning shifts blue.

%%%%%%%%%%%%%%%%%%%%%%%%%%%%%%%%%%%%%%%%%%%%%%%%%%%%%%%%%%%%%%%%%%%
%%%%%%%%%%%%%%%%%%%%%%%%%%%%%%%%%%%%%%%%%%%%%%%%%%%%%%%%%%%%%%%%%%%%%

When considering the coupling between photons and excitons (i.e., $\Omega=10\omega_{m}$ and see Figs. \ref{fig2}(b) and \ref{fig2}(d)), in the absence of pump field ($\varepsilon_{pu}=0$), the probe transmission is suppressed at the red detuning resonant region (black solid line of Fig. \ref{fig2}(b)). It indicates that the attenuated Stokes scattered light is induced. However, the pump field can generates complex transmission spectrum. For a weak pump intensity, there are two attenuated Stokes scattered light beams and one attenuated anti-Stokes scattered light beam (red solid line of Fig. \ref{fig2}(b)). As the pump intensity $\varepsilon_{pu}$ increases, two amplified scattered light beams appear, one is Stokes light and another is anti-Stokes light, while original three attenuated light beams still exist (blue solid line of Fig. \ref{fig2}(b)). As the pump intensity $\varepsilon_{pu}$ further increases, attenuated scattered light beams disappear, while there are only one amplified Stokes light beam and one amplified anti-Stokes light beam, similar to the case without phonon-exciton coupling (see Fig. \ref{fig2}(d)). Furthermore, for two attenuated Stocks scattered light beams, pump fields results in that their amplitudes are both reduced and eventually vanish, while the corresponding resonant probe-pump detuning of left Stokes light shifts red and the one of right Stokes light shift blue. For the attenuated anti-Stocks scattered light beam, the corresponding resonant probe-pump detuning shifts blue as the increase of pump intensity. In a word, the photon-exciton coupling prevent the transmission of probe field signals, this is rooted in that excitons can absorb photons then polaritons are formed, thus output probe field photons are reduced.

%%%%%%%%%%%%%%%%%%%%%%%%%%%%%%%%%%%%%%%%%%%%%%%%%%%%%%%%%%%%%%%%%%%
\begin{figure}[htpb]
\begin{center}
\includegraphics[width=8.0cm]{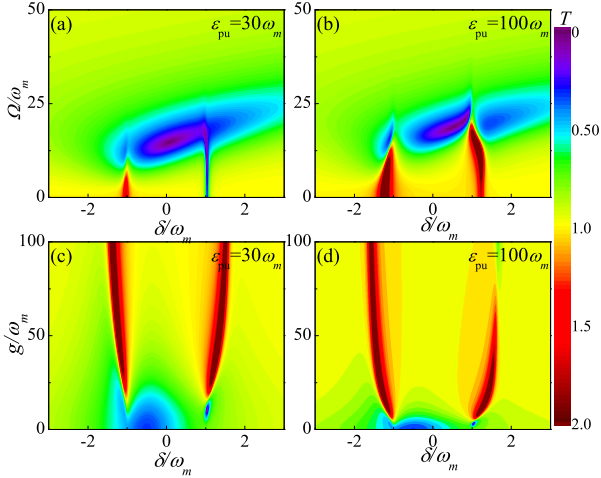}
\end{center}
\caption{Transmission $T$ as function of probe-pump detuning $\delta$ and Rabi coupling strength $\Omega$ (a)-(b) with single-photon optomechanical coupling strength $g=10\omega_{m}$, and as function of $\delta$ and $g$ (c)-(d) with $\Omega=10\omega_{m}$. Other parameters are $\Delta_{a}=\omega_{m}$, $\Delta_{c}=100\omega_{m}$, $\kappa_{a}=\omega_{m}$, $\kappa_{c}=100\omega_{m}$, and $\gamma_{m}=0.1\omega_{m}$.}\label{fig3}
\end{figure}
%%%%%%%%%%%%%%%%%%%%%%%%%%%%%%%%%%%%%%%%%%%%%%%%%%%%%%%%%%%%%%%%%%%%%

The dependence of transmission spectrum on Rabi coupling strength between excitons and photons is more clearly depicted in Figs. \ref{fig3}(a) and \ref{fig3}(b). For a weak pump intensity (Fig. \ref{fig3}(a)), there is an amplified Stokes scattered light near resonant probe-pump detuning $\delta=-\omega_{m}$ and an attenuated anti-Stokes scattered light at $\delta=\omega_{m}$ when photon-exciton coupling $\Omega$ is less than a critical value $\Omega_{c1}$. As $\Omega$ increases ($\Omega<\Omega_{c1}$), the amplitude of the Stokes light is gradually reduced and eventually vanishes, while the amplitude of the anti-Stokes light is gradually reduced, i.e., the transmission of probe field is prohibited near probe-pump detuning $\delta=\omega_{m}$. When photon-exciton coupling $\Omega$ further increases ($\Omega_{c1}<\Omega<\Omega_{c2}$), the probe field transmission is completely prohibited (i.e., $T=0$) near probe-pump detuning $\delta=0$. When photon-exciton coupling $\Omega$ is strong enough (i.e., $\Omega>\Omega_{c2}$), the Stokes and anti-Stokes light beams disappear, and the probe field can almost completely pass through the cavity ($T\sim1.0$) for arbitrary probe-pump detuning, namely, the transmission of probe fields is no longer controlled by pump fields. For a strong pump intensity (Fig. \ref{fig3}(b)), the transmission spectrum is similar to the case of weak pump intensity except that the anti-Stokes scattered light is amplified instead of attenuated for weak photon-exciton coupling.

The transmission spectrum can also be affected by single-photon optomechanical coupling, which is clearly depicted in Figs. \ref{fig3}(c) and \ref{fig3}(d). In the absence of single-photon optomechanical (i.e., $g=0$), there are two attenuated Stokes scattered light beams at blue detuning resonance regions. As single-photon $g$ increases, an attenuated anti-Stokes scattered light beams appears in the red detuning resonance region. As $g$ further increases, Stokes and anti-Stokes light beams are both translated from attenuation to amplification, and the strong pump intensity is conducive to this translation. Therefore, the single-photon optomechanical coupling is beneficial to the amplification of probe field signals.

%%%%%%%%%%%%%%%%%%%%%%%%%%%%%%%%%%%%%%%%%%%%%%%%%%%%%%%%%%%%%%%%%%%
\begin{figure}[htpb]
\begin{center}
\includegraphics[width=8.0cm]{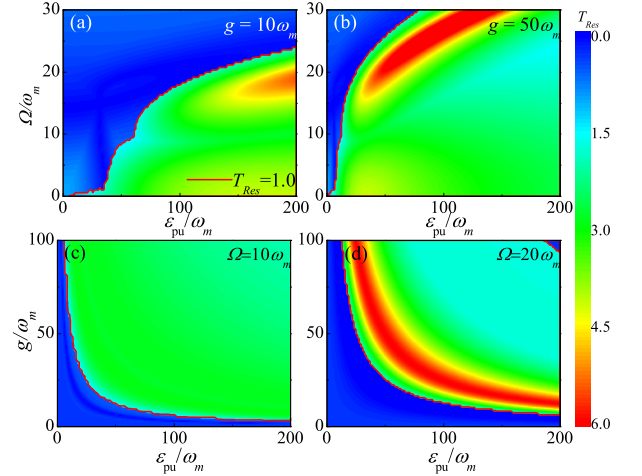}
\end{center}
\caption{The probe laser transmission on the maximum resonance $T_{Res}$ as function of pump laser intensity $\varepsilon_{pu}$ and Rabi splitting $\Omega$ (a)-(b), and as function of pump laser intensity $\varepsilon_{pu}$ and single-photon optomechanical coupling strength $g$. Other parameters are $\Delta_{a}=\omega_{m}$, $\Delta_{c}=100\omega_{m}$, $\kappa_{a}=\omega_{m}$, $\kappa_{c}=100\omega_{m}$, and $\gamma_{m}=0.1\omega_{m}$.}\label{fig4}
\end{figure}
%%%%%%%%%%%%%%%%%%%%%%%%%%%%%%%%%%%%%%%%%%%%%%%%%%%%%%%%%%%%%%%%%%%%%

%%%%%%%%%%%%%%%%%%%%%%%%%%%%%%%%%%%%%%%%%%%%%%%%%%%%%%%%%%%%%%%%%%%
\begin{figure}[htpb]
\begin{center}
\includegraphics[width=8.0cm]{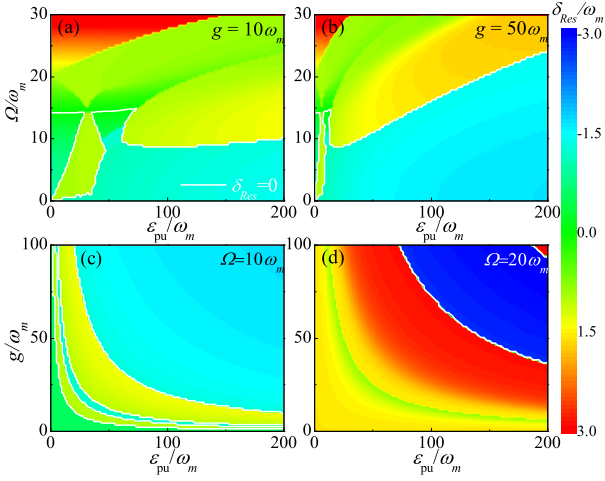}
\end{center}
\caption{The probe-pump detuning on the maximum resonance $\delta_{Res}$ as function of pump laser intensity $\varepsilon_{pu}$ and Rabi splitting $\Omega$ (a)-(b), and as function of pump laser intensity $\varepsilon_{pu}$ and single-photon optomechanical coupling strength $g$ (c)-(d). Other parameters are $\Delta_{a}=\omega_{m}$, $\Delta_{c}=100\omega_{m}$, $\kappa_{a}=\omega_{m}$, $\kappa_{c}=100\omega_{m}$, and $\gamma_{m}=0.1\omega_{m}$.}\label{fig5}
\end{figure}
%%%%%%%%%%%%%%%%%%%%%%%%%%%%%%%%%%%%%%%%%%%%%%%%%%%%%%%%%%%%%%%%%%%%%
\section{Transmission on the maximum resonance}
In the transmission spectrum, there are several resonant regions for a set of input parameters. We focus on the maximum resonance where $|T-1.0|$ have a maximum value, because it can reflect the maximum amplification or attenuation of probe field signals. The corresponding transmission and probe-pump detuning is defined as $T_{Res}$ and $\delta_{Res}$, respectively. On the maximum resonance, probe field signals are amplified for $T_{Res}>1.0$ and attenuated for $T_{Res}<1$. The output light on the maximum resonance is the Stokes scattered light for $\delta_{Res}<0$ and the anti-Stokes light for $\delta_{Res}>0$.

As shown in Fig. \ref{fig4}, on the maximum resonance, the amplification and the attenuation of probe fields depends on pump intensities, photon-exciton Rabi coupling, and sing-photon optomechanical coupling. While pump intensities increase, probe field signals are first translated from attenuation to amplification, then further amplified to a saturation value. When pump intensities unceasingly increase, the transmission on the maximum resonance $T_{Res}$ can be reduced for strong photon-exciton coupling and sing-photon optomechanical coupling (Figs. \ref{fig4}(b) and \ref{fig4}(d)). The photon-exciton Rabi coupling can result in that probe filed signals are translated from amplification to attenuation. These phenomena are because excitons absorb photons to form polaritons, and output photons are reduced. However, for enough strong pump intensities, there is $\Omega_{R1}<\Omega<\Omega_{R2}$, where probe field signals are amplified to a maximum value (Figs.  \ref{fig4}(a) and \ref{fig4}(b)). These phenomena are due to the collective motion of excitons driven by pump fields at resonant probe-pump detuning, which is beneficial to the transmission of photons. The single-photon optomechanical coupling is conducive to the amplification of probe field signals because of the resonance between photons and mechanical resonators. However, for strong photon-exciton coupling (Fig. \ref{fig4}(d)), when probe field transmission reaches the saturation value and single-photon optomechanical coupling further increases, the amplitude of probe field signal amplification decreases and signals are eventually attenuated for enough large single-photon optomechanical coupling and pump intensities. These phenomena are because one mechanical resonator maybe absorb multiple photons for enough strong pump intensities. In the magnified region, probe field signals can be amplified to six times of input laser amplitude at least (i.e., $T_{Res}>6.0$). In the attenuated region, the transmission of probe signals can be inhibited (i.e., $T_{Res}\sim0$). Thus, the probe field signal intensities through such an optomechanical system can be strongly manipulated.

The probe detuning on the maximum resonance $\delta_{Res}$ is demonstrate in Fig. \ref{fig5} to reveal the scattered character of output lasers. The dependence of $T_{Res}$ on Rabi coupling is depicted in Figs. \ref{fig5}(a) and \ref{fig5}(b). In the absence of Rabi coupling (i.e., $\Omega=0$), increasing pump intensities lead to the output scattered light translating from anti-Stokes to Stokes on the maximum resonance (also see Figs. \ref{fig2}(a) and \ref{fig2}(c)). When $\Omega$ increases, increasing pump intensities result in the output scattered light translating from Stokes to anti-Stokes then to Stokes. When $\Omega$ further increases, the output scattered light undergoes a transition from Stokes to anti-Stokes to Stokes then to anti-Stokes. When $\Omega$ is enough large, the output scattered light is always the anti-stokes light. The single-photon optomechanical coupling can enlarge the region of Stokes resonant light (Fig. \ref{fig5}(b)). However, for a weak Rabi coupling (Fig. \ref{fig5}(c)), increasing single-photon optomechanical coupling can lead to output scattered light translating from Stokes to anti-Stokes to Stokes anti-Stokes then to Stokes on the maximum resonance. For a large Rabi coupling (Fig. \ref{fig5}(d)), the output scattered light can undergo a transition from Stokes to anti-Stokes then to Stokes as the single-photon optomechanical coupling increases. Thus, we can accurately control the output probe laser transition between Stokes and anti-Stokes by adjusting related parameters of this optomechanical system.

\section{Conclusion}
In summary, we have proposed a theoretical schema for realizing an all-optical transistor based on a coupled-cavity system with polaritons. A probe laser beam can be modulated by another pump laser beam when it passed throng an optomechanical device which contains cavity photons, excitons, and mechanical oscillator. The output probe laser field modulated by the pump laser field can be either amplified or attenuated owing to exciton-photon coupling and single-photon optomechanical coupling effects. The Stokes and anti-Stokes scattered effect of output prober laser can also be modulated. The discovery could be useful in polariton-based photonic transistors.

\begin{acknowledgments}
This work is supported by the National Key R$\&$D Program of China under Grants
No. 2016YFA0301500, NSFC under grants Nos. 11865014, 11764039, 61775242,
and 61835013, Strategic Priority Research Program of the Chinese Academy of Sciences under grants Nos. XDB01020300, XDB21030300, Natural Science
Foundation of Gansu Province under Grant No. 20JR5RA526, Scientific research project of Gansu higher education under Grand No. 2019A-014, Creation of science and technology of
Northwest Normal University, China, under Grants No. NWNU-LKQN-18-33.
\end{acknowledgments}

\end{document}